\documentclass[pdflatex,sn-mathphys-num]{sn-jnl}

\usepackage{graphicx}%
\usepackage{multirow}%
\usepackage{amsmath,amssymb,amsfonts}%
\usepackage{amsthm}%
\usepackage{mathrsfs}%
\usepackage[title]{appendix}%
\usepackage{xcolor}%
\usepackage{textcomp}%
\usepackage{manyfoot}%
\usepackage{booktabs}%
\usepackage{algorithm}%
\usepackage{algorithmicx}%
\usepackage{algpseudocode}%
\usepackage{listings}%
\usepackage{subcaption}

\theoremstyle{thmstyleone}%
\theoremstyle{thmstyletwo}%

\theoremstyle{thmstylethree}%

\begin{document}

\title{Non-Invasive Reconstruction of Cardiac Activation Dynamics Using Physics-Informed Neural Networks}



\author[1]{\fnm{Nathan} \sur{Dermul}}\email{nathan.dermul@kuleuven.be}

\author[2,3,4]{\fnm{Hans} \sur{Dierckx}} \email{h.j.f.dierckx@lumc.nl}

\affil[1]{\orgdiv{Department of Cardiovascular Sciences, Cardiovascular Imaging and Dynamics}, \orgname{KU Leuven}, \country{Belgium}}

\affil[2]{\orgdiv{Department of Cardiology}, \orgname{Leiden University Medical Centre}, \orgaddress{\street{Albinusdreef 2}, \city{Leiden}, \postcode{2333 ZA}, \country{the Netherlands}}}

\affil[3]{\orgdiv{Mathematical Institute}, \orgname{Leiden University}, \orgaddress{\street{Einsteinweg 55}, \city{Leiden}, \postcode{2333 CC}, \country{the Netherlands}}}

\affil[4]{\orgdiv{Leiden Institute of Physics}, \orgname{Leiden University}, \orgaddress{\street{Niels Bohrweg 2}, \city{Leiden}, \postcode{2333 CA}, \country{the Netherlands}}}


\abstract{

Cardiac arrhythmogenesis is governed by complex electromechanical interactions that are not directly observable in vivo, motivating the development of non-invasive computational approaches for reconstructing three-dimensional activation dynamics. We present a physics-informed neural network framework for recovering cardiac activation patterns, active tension propagation, deformation fields, and hydrostatic pressure from measurable deformation data in simplified left ventricular geometries. Our approach integrates nonlinear anisotropic constitutive modeling, heterogeneous fiber orientation, weak formulations of the governing mechanics, and finite-element-based loss functions to embed physical constraints directly into training.

We demonstrate that the proposed framework accurately reconstructs spatiotemporal activation dynamics under varying levels of measurement noise and reduced spatial resolution, while preserving global propagation patterns and activation timing. By coupling mechanistic modeling with data-driven inference, this method establishes a pathway toward patient-specific, non-invasive reconstruction of cardiac activation, with potential applications in digital phenotyping and computational support for arrhythmia assessment.
}




\maketitle

\section{Introduction}\label{s:introduction}

Cardiac arrhythmias represent a major global health challenge, and arise from complex disturbances in the hearts electrical activity. While atrial fibrillation is the most widespread rhythm disorder, ventricular arrhythmias are particularly dangerous, often leading to sudden cardiac death due to their disruption of the coordinated pumping of blood. Understanding and visualizing the spatiotemporal dynamics of electrical wave propagation is therefore essential for effective diagnosis and treatment. Current clinical practice relies on catheter-based mapping, which provides valuable information but remains invasive, time-consuming, and limited to surface measurements of the myocardium \cite{Zeppenfeld2022}. These constraints highlight the need for alternative non-invasive strategies capable of capturing the full three-dimensional nature of electrical activation. A promising path to such non-invasive assessment of cardiac activation lies in recent advances in three-dimensional echocardiography, which enable rapid and cost-effective measurement of strain-related variables \cite{Lang2018}. By measuring the local onset of mechanical contraction, pilot studies in electromechanical wave imaging (EWI) have identified arrhythmia types and locations. These mechanical markers have been shown, in experimental settings, to reflect the underlying electrical wave propagation in both normal and arrhythmic conditions \cite{Christoph2018}. In contrast to electrocardiographic imaging (EGCi), which relies on projected information on the body surface to visualize wave propagation in the heart, deformation measures provide access to local information \cite{Ramanathan2004,cluitmans2018}. However, measuring mechanical activation is an assessment of the effect of cardiac activation, caused by active tension development, which is in turn triggered by electrical depolarization. In this work, we aim to reconstruct the source of the deformation, i.e. the local active tension developed by the myofibers. 

Previous synthetic studies have already explored classical inversion methods grounded in data assimilation \cite{Beam2020,Kovacheva2021,Lebert2019}. While these techniques are well established in the literature and conceptually robust, their reliance on extensive forward evaluations makes them computationally inefficient when applied to more complex situations with large amounts of degrees of freedom. Alternatively, supervised machine learning approaches have been introduced to solve the inverse problem, first in simple cubic geometries and more recently demonstrated accurate results in handling more complex models in intricate three-dimensional geometries \cite{Christoph2020,Lebert2023}. Once trained, such models can deliver rapid predictions without requiring additional physical input. Yet, the training phase in such methods is computationally intensive, especially without real-world datasets, which then requires extensive simulations. Fully data-driven supervised machine learning frameworks also lack interpretability, require new methodology to insert patient-specific information and may produce results that do not satisfy the laws of physics.\\

Physics-informed neural networks (PINN) strike a balance between classical physics-based inversion methods and the function-approximation and optimization techniques of supervised machine learning\cite{Raissi2019}. By representing functions as neural networks and optimizing them through a loss function that incorporates both data and physics terms, PINNs avoid the need to explicitly solve governing equations \cite{Raissi2019}. In recent years, they have achieved success across multiple fields of applied science \cite{Luo2025}. More specifically, in cardiac electrophysiology PINNs have been used to interpolate electroanatomical mapping data \cite{SahliCostabal2020}, estimate fiber orientations \cite{Magana2025}, and reconstruct tissue parameters \cite{HerreroMartin2022}. Within cardiac mechanics, researchers have utilized these approaches both in forward simulations \cite{Dalton2023} and in the inverse estimation of passive material properties \cite{Caforio2024}. More recently, Caforio et al. have also extended their use to estimating active contractility parameters under homogeneous activation times \cite{Hofler2025}.\\

In earlier work, we demonstrated on synthetic two‑dimensional data that PINNs can be used to infer activation waves, used as a proxy for the underlying electrical activity, from noisy and low resolution deformation data \cite{dermul2024}. In the present study, we extend these results to more complex scenarios, requiring new methodology. A simplified three-dimensional (3D) geometry of a left ventricle (LV) is used, with non‑linear deformation rules as opposed to linear ones, and activation waves are simulated using more realistic biophysically inspired formulations. In addition, strong anisotropy is introduced through spatially varying fiber orientations. 

To address these challenges, we adopt the recently proposed delta‑PINN framework \cite{SahliCostabal2024}. Compared to standard PINNs \cite{Raissi2019}, delta‑PINNs incorporate a transformed input space based on Laplacian eigenfunctions. We optimize with finite‑element‑based losses derived from the weak Galerkin formulation, which was also used in other studies \cite{Gao2022}. These adaptations are well suited to more complex three‑dimensional domains and were shown to facilitate faster convergence and greater accuracy than conventional continuous PINNs or graph neural network approaches. After extending this framework to volumetric data, we estimate 3D active tension wave propagation from deformation data using hyperelastic materials commonly used in cardiac modeling. To satisfy the incompressibility constraint, we simultaneously optimize an auxiliary Lagrange multiplier field $p$, which can be identified with the hydrostatic pressure.\\

This paper is organized as follows. Section \ref{ss:data-generation} describes the generation of the synthetic dataset, including the physics of non‑linear deformation models in cardiac mechanics and the forward simulations. Section \ref{ss:pinn} presents the weak formulation of the governing equations, from which the physics‑based loss term is constructed using finite elements, followed by a discussion of the full PINN optimization framework. Section \ref{s:results} reports the main results: first on the reference dataset, and subsequently on tests assessing the robustness of the PINN under Gaussian noise and reduced spatial resolution. Section \ref{s:discussion} concludes with a discussion of the findings and an outlook on future directions.



\section{Methods}\label{s:methods}

\subsection{Generation of synthetic data}\label{ss:data-generation}

\subsubsection{Physics of active hyperelastic soft tissue}

We work in the framework of non-linear solid mechanics. In the frame of reference of the undeformed material with coordinates $\mathbf{X}$, we then define the new positions of each material point as $\mathbf{x}=\mathbf{x}(\mathbf{X})$ as well as the deformation gradient $\mathbf{F} = \partial\mathbf{x}/\partial\mathbf{X}$, or in index notation: $F_{ij} = \partial x_{i}/\partial X_{j}$. This tensor can be used to transform between frames of reference and $J=\text{det}(\mathbf{F})$ represents the volumetric change. In this study, deformation data $\mathbf{U}$ are related through $\mathbf{F} = \mathbf{I} +\partial\mathbf{U}/\partial\mathbf{X}$. The governing equations of non-linear deformation of an incompressible material in steady-state equilibrium are written in the referential frame as

\begin{align}\label{eq:balance}
    &\nabla\cdot\mathbf{P} = \mathbf{0}\quad\text{in}\:\Omega_0\times [0,T],\\
    &J = 1.\label{eq:balance-2}
\end{align}

The first equation encompasses the equilibrium of stresses inside the material, where $\mathbf{P}$ is the first Piola-Kirchoff stress tensor. The steady-state approximation, meaning there is no acceleration term in the equation, is often assumed in cardiac modeling as the mechanical timescales are much faster than the electrical ones, simplifying the type of partial differential equation (PDE) significantly \cite{Quarteroni2017}. The second equation is the incompressibility constraint, enforcing the material to deform without change in volume. This approximation is also commonly adopted for myocardial tissue because its substantial water content limits volumetric changes. This set of equations has to be closed by suitable boundary conditions.\\

To solve these equations, a constitutive relation must be specified linking deformation to the resulting stresses. Due to the active properties of the myocardium, the stress tensor is first split into a passive $\mathbf{P_p}$ and active part $\mathbf{P_a}$. The passive contribution is modeled by hyperelastic material which are defined by a well chosen energy-density function $\mathcal{W}$ such that $\mathbf{P_p}=\partial \mathcal{W}/\partial \mathbf{F}$. In this work, we make use of the Guccione model \cite{Guccione1995}

\begin{align}\label{eq:guccione} 
    &\mathcal{W}_{guccione} = \frac{C}{2}(e^Q-1)\\
    &Q = b_fE_{F,11}^2 + b_t(E_{F,22}^2 + E_{F,33}^2 + 2E_{F,23}^2) + b_{fs}(2E_{F,12}^2 + 2E_{F,13}^2)
\end{align}

with 4 scalar parameters $C, b_f, b_t, b_{fs}$. $\mathbf{E_{F}}$ denotes the Green-Lagrange strain tensor $\mathbf{E} = \frac{1}{2}(\mathbf{F}^T\mathbf{F}-\mathbf{I})$, rotated in the local fiber-sheet-normal coordinate system, i.e., $\mathbf{E_{F}} = \mathbf{Q}^T\mathbf{E}\mathbf{Q}$ with $\mathbf{Q} = [\mathbf{f_0},\mathbf{s_0},\mathbf{n_0}]$. This results in axially symmetric, anisotropic response of the passive tissue. The corresponding stress tensor can then be written as \cite{Hadjicharalambous2015a}

\begin{equation}\label{eq:p_p}
    \mathbf{P_p} = \frac{\partial W_{guccione}}{\partial \mathbf{F}}=\mathbf{F}Ce^Q\mathbf{Q}(\mathbf{a}\circ\mathbf{E_F} )\mathbf{Q}^T
\end{equation}

where the matrix $\mathbf{a}$ includes the anisotropy values

\begin{equation}
    \mathbf{a} = \begin{bmatrix}
b_f & b_{fs} & b_{fs}\\
b_{fs}  & b_t& b_t\\
b_{fs}  & b_t & b_t
\end{bmatrix}.
\end{equation}

The active contribution to the stress is more directly formulated as

\begin{equation}\label{eq:p_a}
    \mathbf{P_a} ={T_a}\mathbf{F}\cdot (\mathbf{f_0} \otimes \mathbf{f_0}).
\end{equation}

where the generated force acts only in the direction of the fibers. Finally, to enforce the incompressibility constraint (Eq. \ref{eq:balance-2}), an additional term $\mathcal{W}_{vol}= -p(J-1)$ is added to the passive contribution, penalizing volumetric changes. The factor $p$ serves as a Lagrange multiplier and can be identified with the hydrostatic pressure. We thus end up with the following total stress tensor

\begin{equation}\label{eq:p_tot}
    \mathbf{P} = \mathbf{P_p}  + \mathbf{P_a} - pJ\mathbf{F}^{-T}.
\end{equation}

\subsubsection{Forward simulations}

To evaluate our inversion method, synthetic data were generated by solving the full cardiac electro-mechanical equations. Simulations were performed using the publicly available software package \textit{simcardems} \cite{Finsberg2023}, a FEniCS-based Finite Element solver written in python to simulate cardiac electro-mechanics, developed and maintained by the simula research group. The reference configuration of the undeformed left ventricle (LV) was represented by an idealized three-dimensional ellipsoid, generated with the software package \textit{cardiac-geometries} \cite{finsberg2022}, similar to other computational studies \cite{Land2015}. The LV base was modeled as a circle of radius $10$\;mm, centered at the point $(0,\ 0,\ 5)$, with the apex located at $z=-20$\;mm. The transmural wall thickness was set to $3$\;mm. Additionally, rule-based fibers were created, varying from $60^o$ on the endocardial to $-60^o$ on the epicardial surface. The final mesh contained 19061 tetrahedral elements connecting 4708 nodes, with two to three elements across the wall thickness.\\

The electrical wave propagation was modeled by the monodomain formulation of electrophysiology, while the local cell dynamics were described by the O'Hara-Rudy model, coupled to the Land model to generate active tension \cite{OHara2011,Land2017}. Default parameter values for both models were used, see the simcardems GitHub documentation. The computed $T_a$ was then provided as input to the mechanics solver at each timestep. The quasi-static balance equations of non-linear deformation (Eq. \ref{eq:balance}) were solved in the finite element (FEM) framework, while the incompressibility constraint was enforced by employing the Lagrange multiplier method in which one solves for both the displacement $\mathbf{U}$ and the hydrostatic pressure $p$ simultaneously. This mixed FE formulation used Taylor-Hood elements consisting of quadratic Lagrangian elements for $\mathbf{U}$ and linear elements for $p$. The passive stresses were modeled using the Guccione constitutive law (Eq. \ref{eq:guccione}) with parameters $C=2$\:kPa, $b_f=8$, $b_t=2$, $b_{fs}=4$, as in \cite{Land2015} and assumed to be known in this study. Mechanical boundary conditions were imposed by constraining the LV base with no-deformation Dirichlet conditions, while the endocardial and epicardial surfaces were assigned traction-free Neumann conditions.\\

The simulation was initiated by injecting a short current stimulus at the endocardial position $(0,7,-5)$ within a sphere of radius $2$\;mm. This triggered the propagation of the electrical wave, followed by the delayed rise of intracellular calcium and the subsequent development of active tension. The final dataset used in this work, consisted of nodal values for the active tension $T_a(\mathbf{r},t)$, mechanical deformation $\mathbf{U}(\mathbf{r},t)$ and hydrostatic pressures $p(\mathbf{r},t)$ recorded over a duration of $100$\;ms with an output interval of $2$\;ms. It should be noted that the simulation itself employed finer time steps of $0.5$\;ms and the electrophysiology was solved on a refined mesh, necessary to ensure convergent solutions.

\subsubsection{Postprocessing}

To test the robustness of our method, we reduced the quality of deformation data by either introducing additive noise or lowering the spatial resolution. Gaussian noise was added to each component $j$ of the deformation vector such that new data points $U_j$ were given by

\begin{equation}
    U_j = U_{j,0} + \eta_j\;p\;U_{j,max}
\end{equation}

where $U_{j,0}$ are the original values and $U_{j,max}$ is the maximum value over the full spatial and temporal range. Moreover, $\eta_j$ were sampled from a standard normal distribution, while $p$ is the noise level of choice, expressed as a fraction of the maximum deformation.\\

To mimic loss of spatial resolution, while retaining the original volumetric mesh, we used a combination of smoothing and subsampling. First, all components of the deformation were smoothed with a Gaussian filter with standard deviation equal to $\sigma$, expressed in units of average vertex length $\approx 1.16$\;mm. The discrete filter used all data points in a radius of $4\sigma$. Second, $s = 1/8^{sigma}$ data points were randomly sampled from the original dataset. This reflects the principle that halving the resolution in all three dimensions reduces the number of retained points to one eighth, effectively averaging the information over one characteristic length i.e $\sigma=1$.

\subsection{Optimization of the PINN}\label{ss:pinn}

We adopted PINNs to estimate underlying active tension waves from the resulting deformation. These methods combine the performance of neural networks  in data fitting with ability to work with sparse data and impose physical or physiological constraints. In general, PINNs can be seen as flexible nonlinear function approximators, optimized on the combination of available data and physical equations. Here, we utilize the recently proposed Delta-PINN framework \cite{SahliCostabal2024}, which was introduced to handle more complex spatial domains as well as to facilitate the learning process. First, we discuss the formulation and numerical calculation of the incorporated physics and afterwards present the full optimization scheme of the PINN.\\

\subsubsection{Weak formulation of the mechanical equations}

The physics information is enforced based upon a weak formulation of the governing equations. This lowers the order of derivative and incorporates Neumann boundary conditions naturally. The weak formulation of the balance equation (Eq. (\ref{eq:balance})) is obtained with the standard Galerkin variational methodology, where the strong formulation is multiplied with a test vector function $\mathbf{v}(\mathbf{r})$ and integrated over the reference domain $\Omega_{0}$:

\begin{equation}
    \int_{\Omega_0}(\nabla\cdot\mathbf{P})\cdot\mathbf{v}dV = 0.
\end{equation}

The divergence theorem then allows splitting the integral into two parts 

\begin{equation}
    \int_{\partial\Omega_0}(\mathbf{P}\cdot\mathbf{v})\cdot\mathbf{n}dS - \int_{\Omega_0}(\mathbf{P}:\nabla\mathbf{v})dV = 0.
\end{equation}

Next, at every boundary surface $\partial\Omega_0$, a Dirichlet or Neumann boundary condition is imposed. For the latter we can write $\mathbf{P}\cdot\mathbf{n} = \mathbf{t}$ which is the traction vector on the boundary and is connected to the pressures on boundary surfaces such as the endo-or epicardium. In this work, these are set to zero. The former will also vanish if the test function is chosen to be zero at the Dirichlet boundary. We thus end up with following integral over the whole reference domain

\begin{equation}\label{eq:weak-formulation}
    \int_{\Omega_0}(\mathbf{P}:\nabla\mathbf{v})dV = 0
\end{equation}

which has to be satisfied for all test functions $\mathbf{v}(\mathbf{r})$.\\

\subsubsection{Finite element formulation}

To calculate the residual (Eq. \ref{eq:weak-formulation}) numerically, all fields of interest are represented by finite elements formulated on a 3D volumetric mesh. We thus need only to estimate nodal values, which reduces the search space in comparison to the optimization of continuous functions in classical PINNs. This approach also allows to enforce Dirichlet boundary conditions in a hard way. To lower the computational cost of our method, we make use of linear tetrahedral elements with $N$ shape functions $N_k(\mathbf{r})$ for each node $k$. The test function $\mathbf{v}(\mathbf{r})$ (similar to the trial function $\mathbf{u}(\mathbf{r})$)is then calculated as the multiplication of the (N,1) vector of shape functions and the 2nd order tensor $\mathbf{\hat{v}}$ with shape (N,3) representing the nodal values of all three components

\begin{equation}
    \mathbf{v}(\mathbf{r}) = \mathbf{N}(\mathbf{r})\cdot\mathbf{\hat{v}} = \begin{bmatrix}
            N_{1}(\mathbf{r}) & N_{2}(\mathbf{r}) & N_{3}(\mathbf{r}) & ...
                \end{bmatrix} \cdot \begin{bmatrix}
v_{1,x} & v_{1,y} & v_{1,z}\\
v_{2,x} & v_{2,y} & v_{2,z}\\
v_{3,x} & v_{3,y} & v_{3,z}\\
... & ... & ... 
\end{bmatrix}.
\end{equation}

We can then insert this in our variational residual formulation (Eq. \ref{eq:weak-formulation}) to find following expression in index notation

\begin{equation}
     \int_{\Omega_0}\Sigma_{i}^{3}\Sigma_{j}^{3}P_{ij}(\mathbf{r})\partial_{i}(\Sigma_{k}^{N}N_k(\mathbf{r})v_{kj})dV = 0.
\end{equation}

This equation should hold for any test function (that vanishes on the Dirichlet boundary), so for any values of $v_{kj}$. This leads to the tensor equation

\begin{equation}
      \Sigma_{i}^{3}\int_{\Omega_0}P_{ij}\partial_{i}(N_k)dV = 0
\end{equation}

for every $j=1,2,3$ and $k=1,2,3,..N'$ with $N'= N - N_{dir}$ taking into account the $N_{dir}$ Dirichlet boundary nodes. To calculate the integral we make use of the finite element $N_k$ being only non-zero in elements neighboring the node $k$. Additionally, working with linear elements for the deformation $\mathbf{U}$, active tension $T_a$, hydrostatic pressure $p$ as well as (approximately) linear fiber direction data, we can see from Eqs. (\ref{eq:p_p}-\ref{eq:p_tot}) that the active part of $\mathbf{P}$ will be of cubic order, while the pressure term is linear and the Guccione contribution exponential. For this reason we calculate the integrals over the elements with Gaussian quadrature rules which are exact up to order 3. With the 5 quadrature points $\mathbf{r}_q$ inside one element and corresponding weights $w_q$, we can write the residual then as

\begin{equation}\label{eq:weak-residual}
      R_{j,k} = \Sigma_{i}^{3}\Sigma_{e}^{e_k}\Sigma_{q}^{5}(P_{ij}(\mathbf{r}_q))_e\;(\partial_{i}(N_k))_{e} \; V_e = 0
\end{equation}

for every $j=1,2,3$ and $k=1,2,3,..N'$, while using $e_k$ to denote all elements which include node k.\\

\subsubsection{Optimization Problem}

\begin{figure}[ht!]
    \centering
    \includegraphics[width=1.0\textwidth]{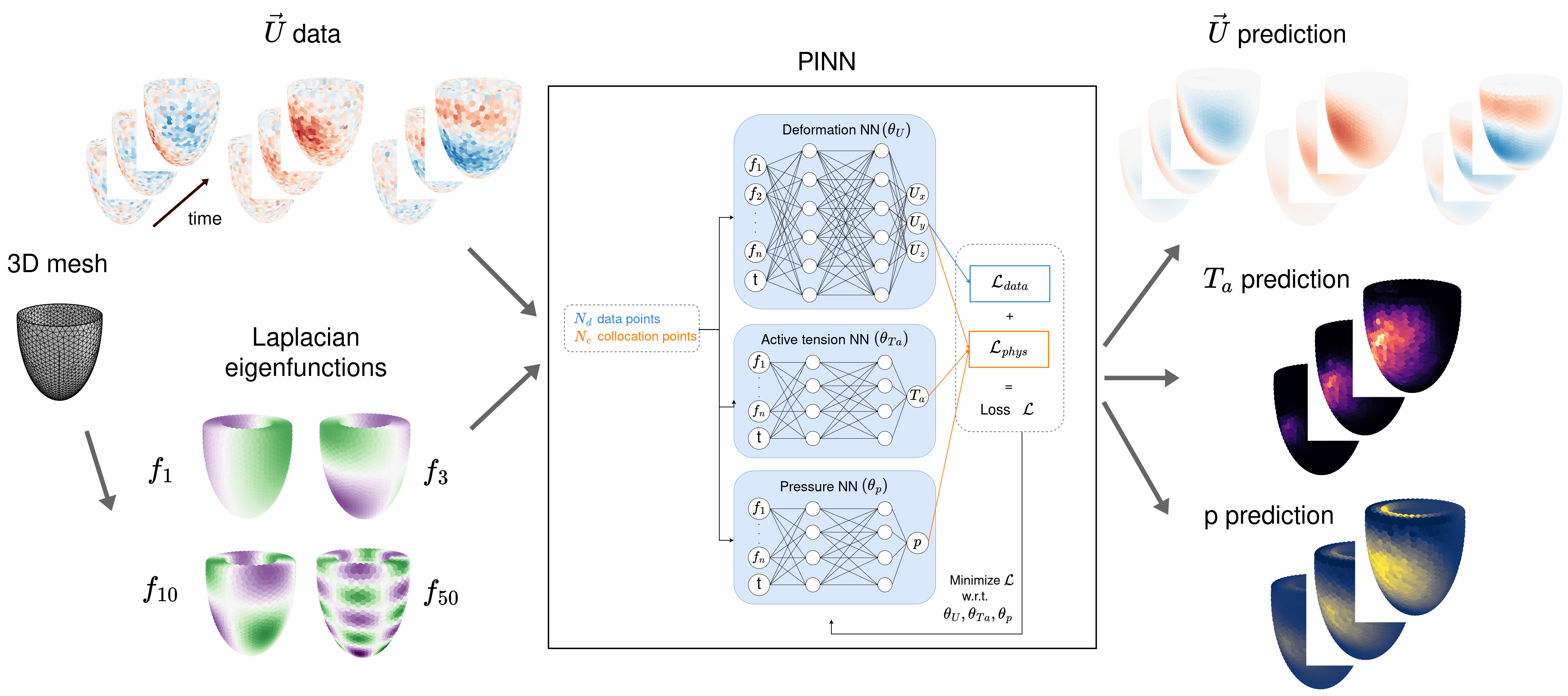}
    \caption[Schematic overview of 3D PINN method]{Schematic overview of the PINN framework. The input consists of a 3D volumetric mesh together with available deformation data defined on the vertices. Laplacian eigenfunctions $f_{n}(\mathbf{r})$ are then calculated as a preprocessing step and serve as the input coordinates for the PINN. The spatial-temporal fields of the deformation vector $\mathbf{U}(\mathbf{r},t)$, active tension $T_a(\mathbf{r},t)$ and hydrostatic pressure $p(\mathbf{r},t)$ are represented by three fully connected NNs. The optimization is done by minimizing two terms that make up the total loss function: the data loss which is evaluated on $N_d$ data points and the physics loss enforced on $N_c$ collocation points.}
    \label{fig:scheme_pinn}
\end{figure}

A schematic overview of the PINN methodology to estimate the spatiotemporal active tension field is shown in Fig. (\ref{fig:scheme_pinn}). The input consists of a 3D volumetric mesh and available deformation data $\mathbf{U}_i$, defined on the nodes of the mesh. An key feature of the Delta-PINN framework \cite{SahliCostabal2024} is to employ the eigenfunctions of the Laplacian operator as the spatial input space of the PINN. In comparison to the spatial coordinates of the original PINNs \cite{Raissi2019,SahliCostabal2020,dermul2024}, they naturally incorporate the 3D domain information in the optimization problem and help to facilitate learning by providing more suitable and flexible input features. The calculation of the eigenfunctions is done before the optimization as a preprocessing step by solving the corresponding eigenvalue problem numerically on the tetrahedral mesh.\\

We represent our fields of interest, the deformation vector $\mathbf{U}(\mathbf{r},t)$, the active tension $T_a(\mathbf{r},t)$ and the hydrostatic pressure $p(\mathbf{r},t)$ as separate fully connected Neural Networks (NN) with parameters $\theta_{U}$, $\theta_{T_a}$ and $\theta_{p}$. The input nodes of the NN consist of the first n eigenfunctions $f_{n}(\mathbf{r})$ together with the time coordinate $t$, while the fields of interest form the output nodes. We chose here to combine the deformation components in one NN as these are generally connected and can share similar patterns. The active tension field is first transformed through a sigmoid function and multiplied by $T_{max} = 10$\;kPa to only allow positive values. In addition, the time coordinates and deformation outputs are standardized with mean $0$ and std $1$, which facilitates the optimization of the NN.\\ 

The optimization of the PINN is done by minimizing a loss function $\mathcal{L}$ defined as the weighted sum of two different terms

\begin{equation}
    \mathcal{L} = \mathcal{L}_{data} + \alpha\mathcal{L}_{phys}
\end{equation}

for N iterations. During every iteration, $N_d$ data points are randomly selected and the data loss term is calculated in the standard quadratic manner as 

\begin{equation}\label{eq:total-loss}
    \mathcal{L}_{data} = \frac{1}{N_d}\sum_{d=1}^{N_d}||\hat{\mathbf{U}}_d-\mathbf{U}(\mathbf{r}_d,t_d)||^2
\end{equation}

 where $\hat{\mathbf{U}}_d$ is the deformation vector the selected data points and $\mathbf{U}(\mathbf{r}_d,t_d)$ are its current predictions. To enforce the momentum balance equation on the solution, we need to minimize the residual from Eq. (\ref{eq:weak-residual}). As a matrix equation, it needs to be fulfilled for all directions $j=1,2,3$ and all nodes $k=1,2,3,..N$. Similar to the data term, we can formulate this in mini-batches as following. First, $N_c$ collocation nodes are randomly sampled over the full spatial-temporal domain, and all relevant properties for element which include the sampled nodes are gathered as well. This constitutes of variables which are constant over an element i.e. the volume and gradient of the shape functions and variables which need to be known at the quadrature points namely the spatial coordinates (eigenfunction values) and the fiber directions. Note that all these matrices are precomputed. Then, for every sampled node $k$ we calculate all three $j$ components of the residual Eq. (\ref{eq:weak-residual}). The terms will depend on the current predicted fields at the quadrature points. Finally, if we denote all necessary evaluations of variable $Z$ i.e. at all quadrature points in elements around the sampled node k, as $Z(\mathcal{N}(k)) = Z(\mathbf{r}_{\mathcal{N}(k)},t_{\mathcal{N}(k)})$, we can formulate the physics loss function as

 \begin{equation}
    \mathcal{L}_{phys} = \frac{1}{N_c}\sum_{k}^{N_c}\sum^{3}_{j}R_{j,k}(\mathbf{U}(\mathcal{N}(k)), T_a(\mathcal{N}(k)), p(\mathcal{N}(k)))^2.
 \end{equation}

Note that all nodes $k$ can be chosen except those with Dirichlet boundary conditions. Moreover, because the boundary conditions are strictly enforced, Dirichlet boundary nodes are assigned zero values when evaluating the physics term, thereby removing those degrees of freedom from the system. Before the $N$ iterations on the full loss function, we first optimize the PINN with only the data term (i.e. $\alpha=0$) for $2000$ iterations. This step ensures that the initial deformations do not lead to excessively large values in the physics loss, which depends exponentially on them. In both phases, the optimization is done using the Adam optimizer \cite{Kingma2014AdamAM} and its default parameters, while all NNs are initialized according to the Glorot scheme.\\

Following hyperparameters were kept constant for all experiments. We chose to include the first $200$ eigenfunctions as these expressed a good range of low and higher frequency signals and the necessary the necessary variation across the myocardial wall (starting from around $100$), without making the optimization overly challenging by introducing too much redundant inputs. Similarly, hidden layers of $[50,50,25]$ were sufficient to represent the type of data patterns in this study. Mini-batches contained $N_d=N_c=64$ nodal points and the number of iterations $N$ was set to $60000$ ($N_{pre}=2000$) where both loss terms were seen to be converged. To select the weights $\alpha$, initial values of $10^0$ were tested and restarted after $5000$ iterations with one order of magnitude lower when the loss exhibited unstable behavior. All optimizations were executed on a HP Z-book, 11th Gen Intel Core i7-11800H, while JAX's \cite{jax2018github} automatic CPU parallelization used all 16 available virtual nodes, taking on average $3$\:h of computation time. As a postprocessing step, we also calculated the local activation time (LAT) of the active tension wave by identifying, for each spatial point, the first timestep at which $T_a$ exceeded the threshold of $0.5$\;kPa.

\section{Results}\label{s:results}

All results shown were obtained by utilizing the constant hyperparameters mentioned in the previous section. The selected $\alpha$ values for each dataset can be seen in Table \ref{tab:rmse}. Note that in noisy data sets, the data loss relatively high, and the physics term can get larger. Contrarily, when less data points available (e.g. with reduced resolution) this is the opposite.

\subsection{Reference solution}

\begin{figure}[h!]
    \centering
    
    \begin{subfigure}{0.4775\textwidth}
        \includegraphics[width=\linewidth]{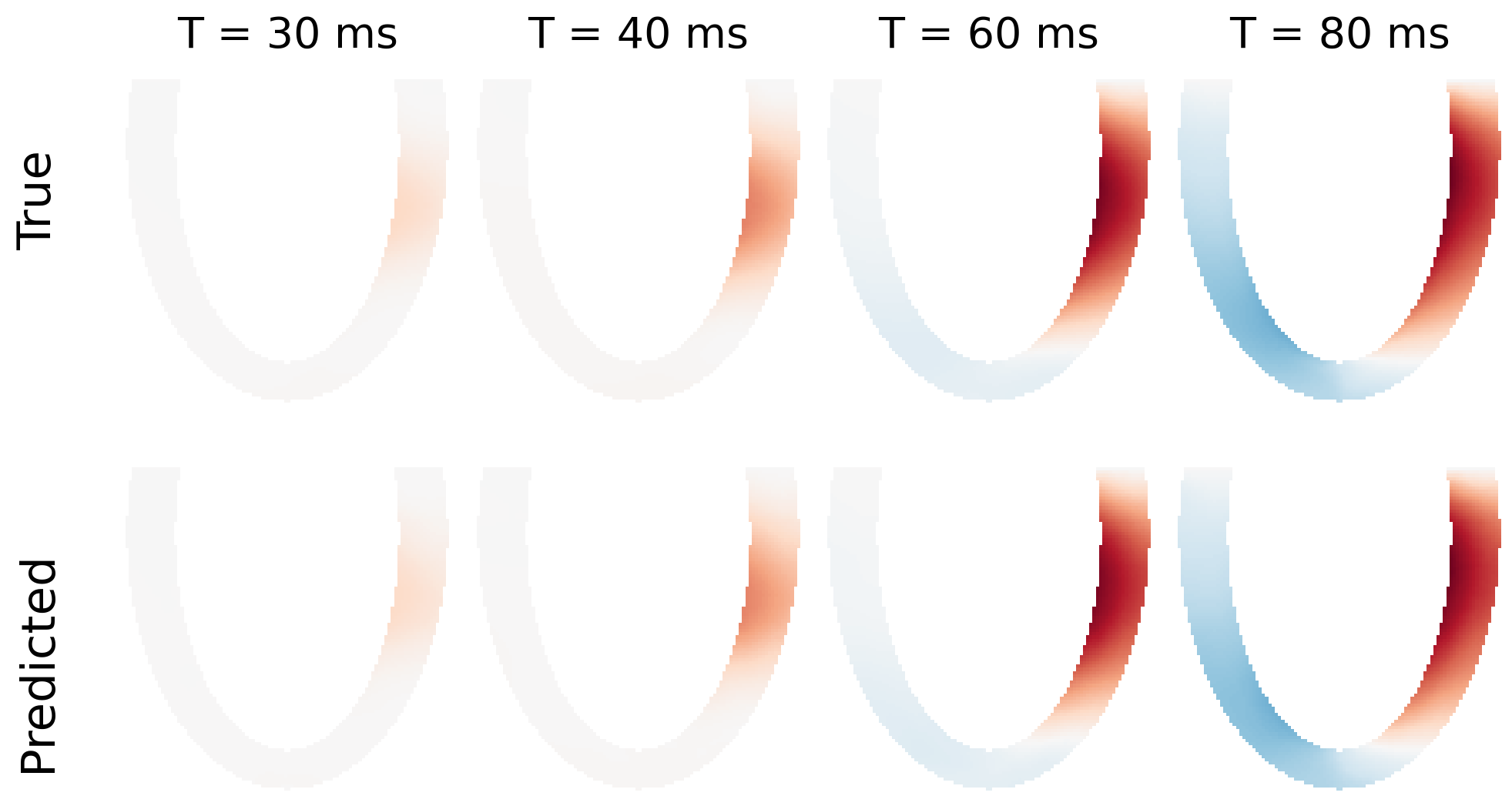}
        \caption{}
        \label{fig:reference-u-hp-a}
    \end{subfigure}
    \hfill
    \begin{subfigure}{0.5025\textwidth}
        \includegraphics[width=\linewidth]{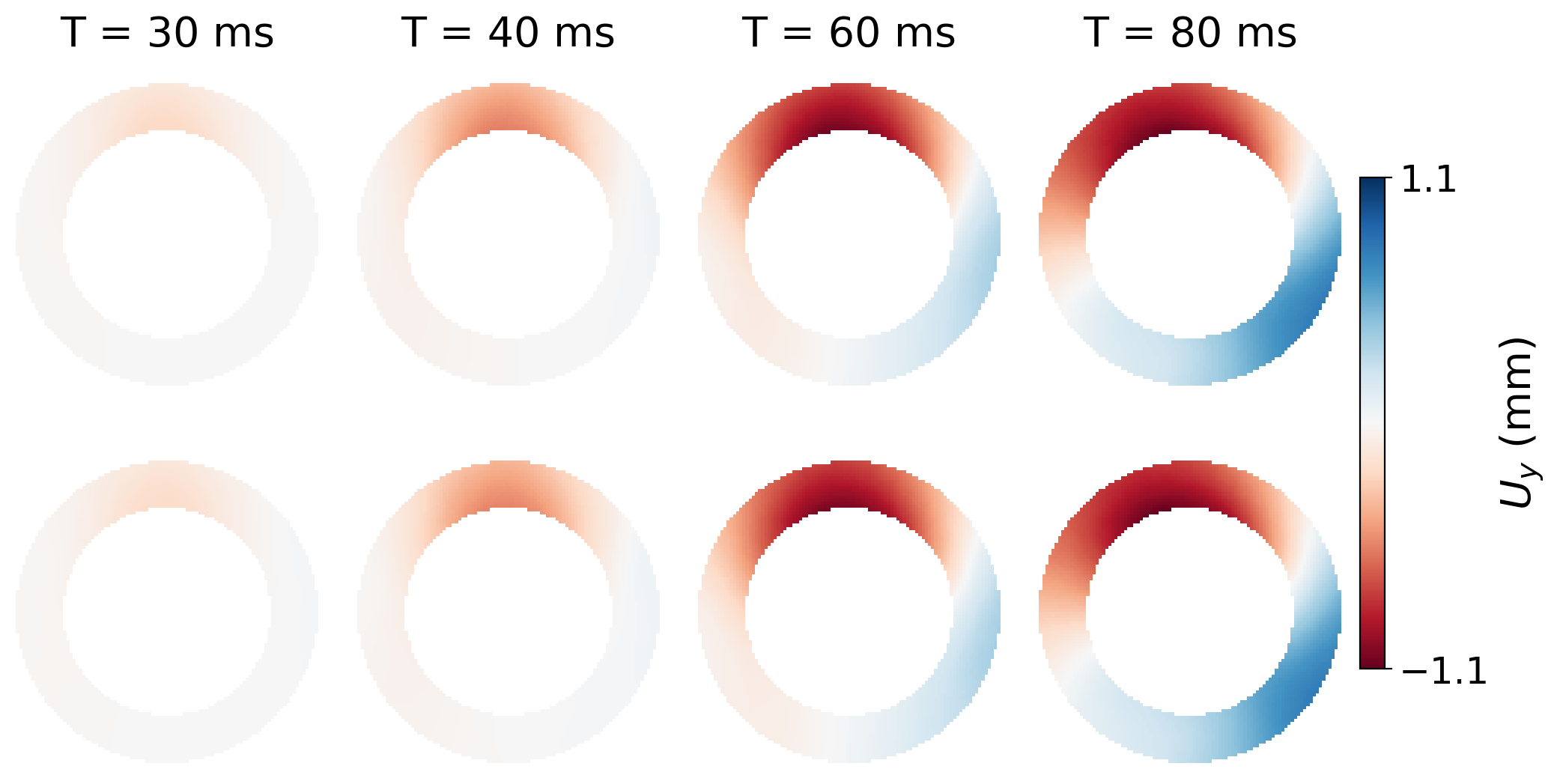}
        \caption{}
        \label{fig:reference-u-hp-b}
    \end{subfigure}
    
    \begin{subfigure}{0.4775\textwidth}
        \includegraphics[width=\linewidth]{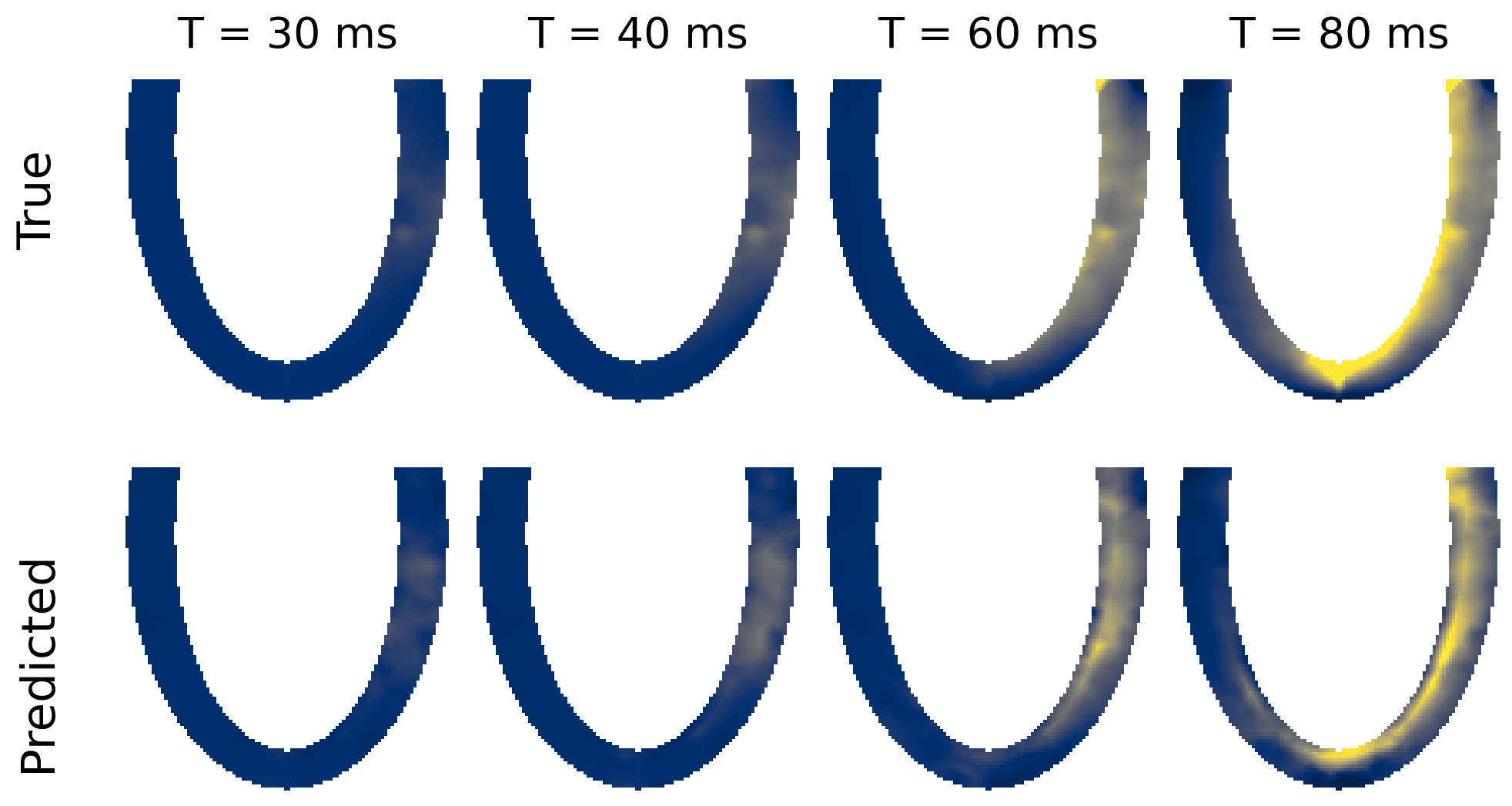}
        \caption{}
        \label{fig:reference-u-hp-c}
    \end{subfigure}
    \hfill
    \begin{subfigure}{0.5025\textwidth}
        \includegraphics[width=\linewidth]{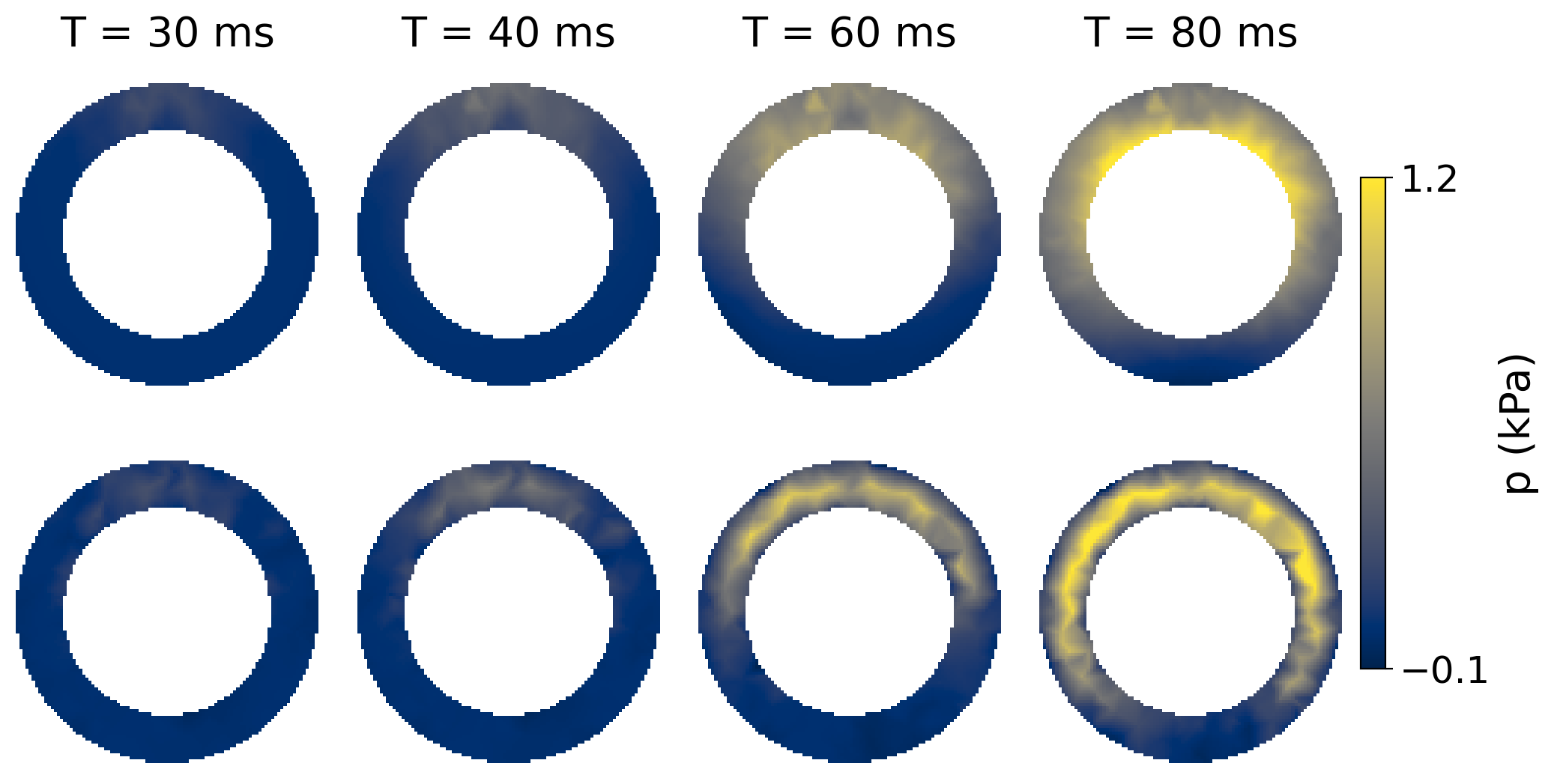}
        \caption{}
        \label{fig:reference-u-hp-d}
    \end{subfigure}
    
    \caption[3D PINN results ($\mathbf{U}$ and $p$) for reference data]{PINN results for the estimated deformation field $U$ and hydrostatic pressure $p$, optimized on all available data points. The fields are visualized by slicing the 3D volumetric mesh with the $x=0$\;mm plane (longitudinal, left) and the $z=-5$\;mm plane (radial, right). Both planes slice through the source point, see Fig. \ref{fig:noise} for the 3D orientation. Panels \textbf{(a)} and \textbf{(b)} show the y-component of the deformation, while panels \textbf{(c)} and \textbf{(d)} consist of the hydrostatic pressure $p$, used in the forward simulation to satisfy the incompressibility constraint.}
    \label{fig:reference-u-hp}

\end{figure}

\begin{figure}[h!]
    \centering
    
    \begin{subfigure}{0.9\textwidth}
        \includegraphics[width=\linewidth]{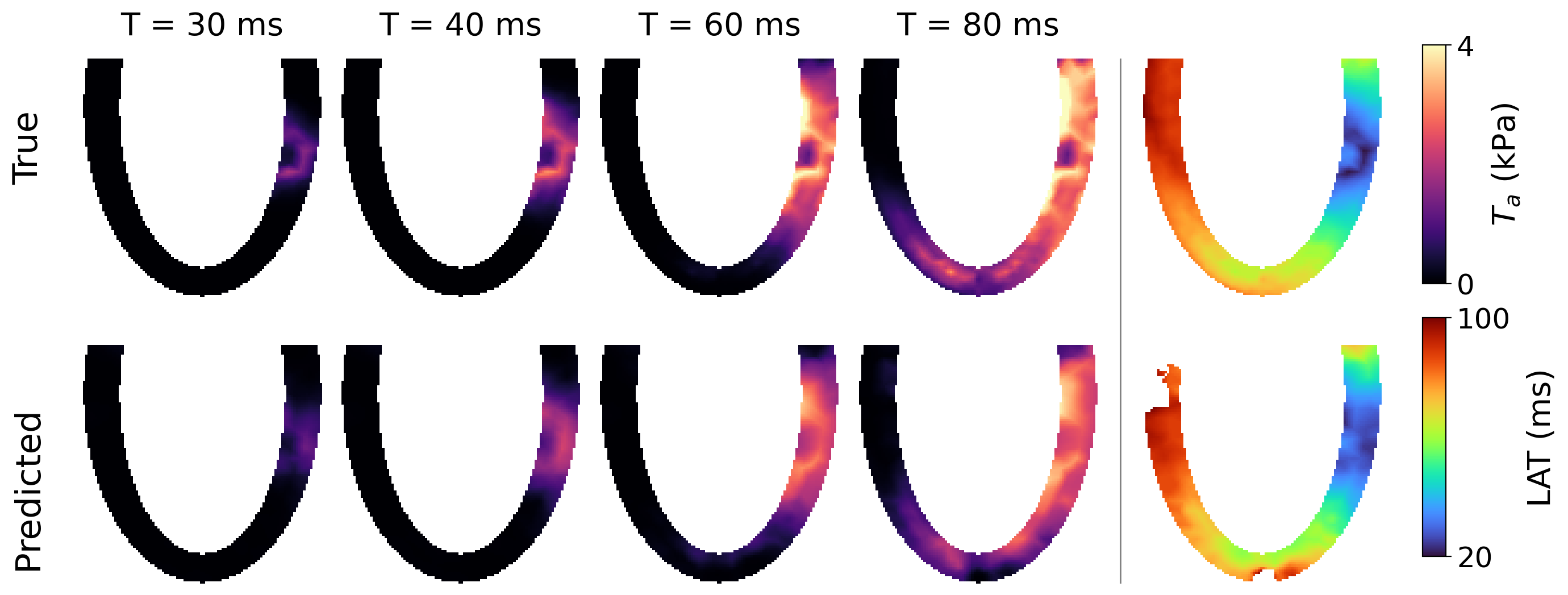}
        \caption{}
        \label{fig:reference-Ta-a}
    \end{subfigure}
    
    \begin{subfigure}{0.9\textwidth}
        \includegraphics[width=\linewidth]{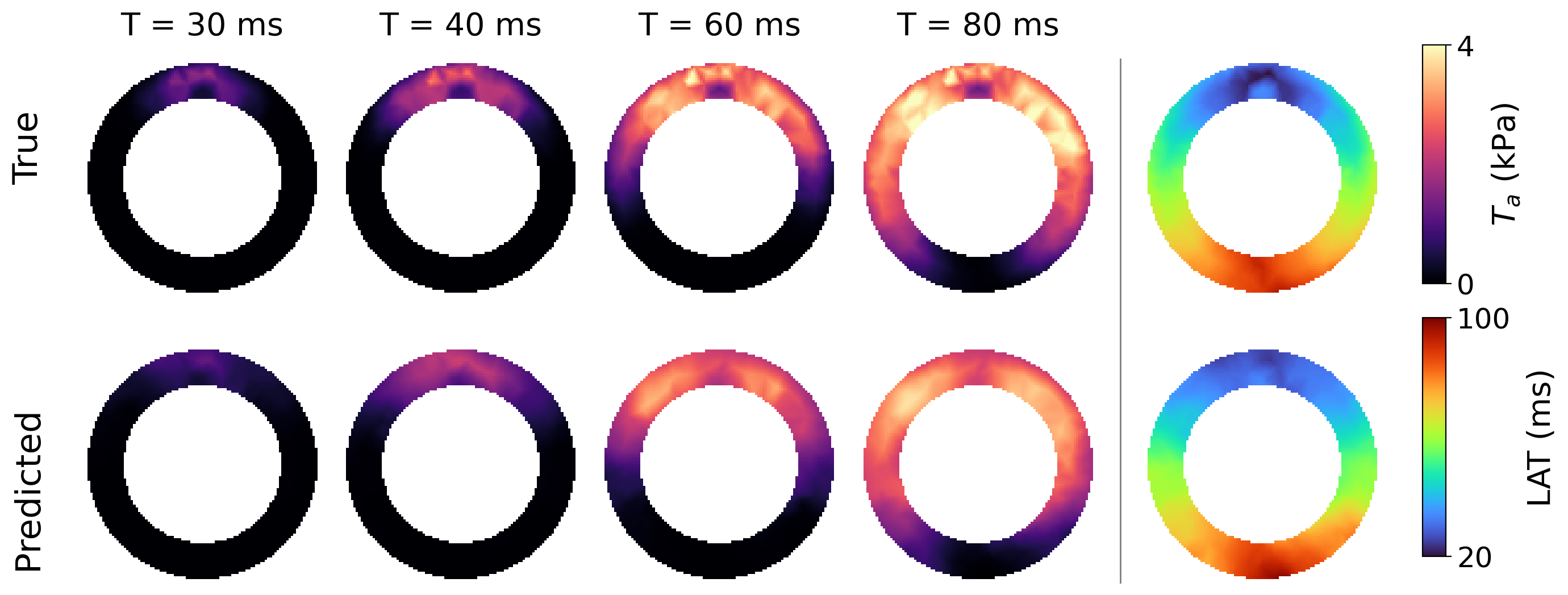}
        \caption{}
        \label{fig:reference-Ta-b}
    \end{subfigure}
    
    \caption[3D PINN results ($T_a$) for reference data]{PINN results for the estimated active tension $T_a$, optimized on all available data points. The fields are visualized by slicing the 3D volumetric mesh with the $x=0$\;mm plane (longitudinal, panel \textbf{(a)}) and the $z=-5$\;mm plane (radial, panel \textbf{(b)}). Both planes slice through the source point, see Fig. \ref{fig:noise} for the 3D orientation. In addition to the $T_a$-field, the last column shows the corresponding LAT map for the activation wave.}
    \label{fig:reference-Ta}

\end{figure}

In this section, we present the PINN results for the reference simulation data, i.e. when no Gaussian noise added and all data points are included. Figs. \ref{fig:reference-u-hp} and \ref{fig:reference-Ta} show the predicted fields in two distinct slices throughout the 3D volumetric mesh evaluated at different times during wave propagation. Both slices intersect the source location: the longitudinal view corresponds to the plane at  $x=0$\;mm, while the radial view corresponds to the $z=-5$\;mm plane, see the first column of Figs. \ref{fig:noise} and \ref{fig:resolution} for the 3D orientation. All fields are visualized by the underlying linear finite elements. 

Panels (a) and (b) of Fig. \ref{fig:reference-u-hp} show the y-component of the deformation field as a representative component. The PINN is able to accurately learn the deformation data points. The overall evolution over time and finer differences between endo- and epicardial surfaces are both recovered in the PINN predictions. Similar results are obtained for the other deformation components. Panels (c) and (d) compare the true to the estimated Lagrange multiplier $p$ (pressure) used in the forward simulation to satisfy the incompressibility constraint and identified with the hydrostatic pressure of the blood. Again, the global evolution through the ellipsoid is captured in the PINN and we see comparable magnitudes for every timestep. However, the PINN prediction seems to avoid large values on the surfaces, especially on the endocardium and the estimation is also less smooth than the true field. This fragmented behavior could be related to the indirect violation of the inf-sup condition in mixed FE formulations in which the Lagrange field ($p$) should be one order less than the main field of interest i.e. the deformation $\mathbf{U}$ \cite{Gatica2014}.\\

The main objective of this study however, lies in the estimation of the active tension field which represents the underlying signals contracting the tissue. Fig. \ref{fig:reference-Ta} shows the PINN predictions of the active tension $T_a$ in the same 2D-slices as the previous Fig. \ref{fig:reference-u-hp}. The last column consists of the LAT maps, calculated after the optimization as described earlier. The active tension wave is reconstructed accurately, starting from the source position at $T=30$\;ms, while propagating and increasing in magnitude throughout the ellipsoid. This is also reflected in the corresponding LAT maps which clearly show the initial onset and the subsequent progression to the other side, resulting in a symmetrical map in panel (b) due to the specific slice orientation. Additionally, we observe that finer details in the $T_a$-field are partially recovered by the PINN as well, such as the absence of $T_a$ at the source point and the smaller magnitudes at the apex and at the epicardial surface. While the first of these properties appears to be an artificial consequence of the numerical forward simulation, it is nevertheless encouraging that the PINN can reconstruct the $T_a$‑field on finer spatial scales, despite the inherent data–model mismatch in the function space of $\mathbf{U}$.\\ 

\begin{figure}[h!]
    \centering
    
    \begin{subfigure}{1.0\textwidth}
        \includegraphics[width=\linewidth]{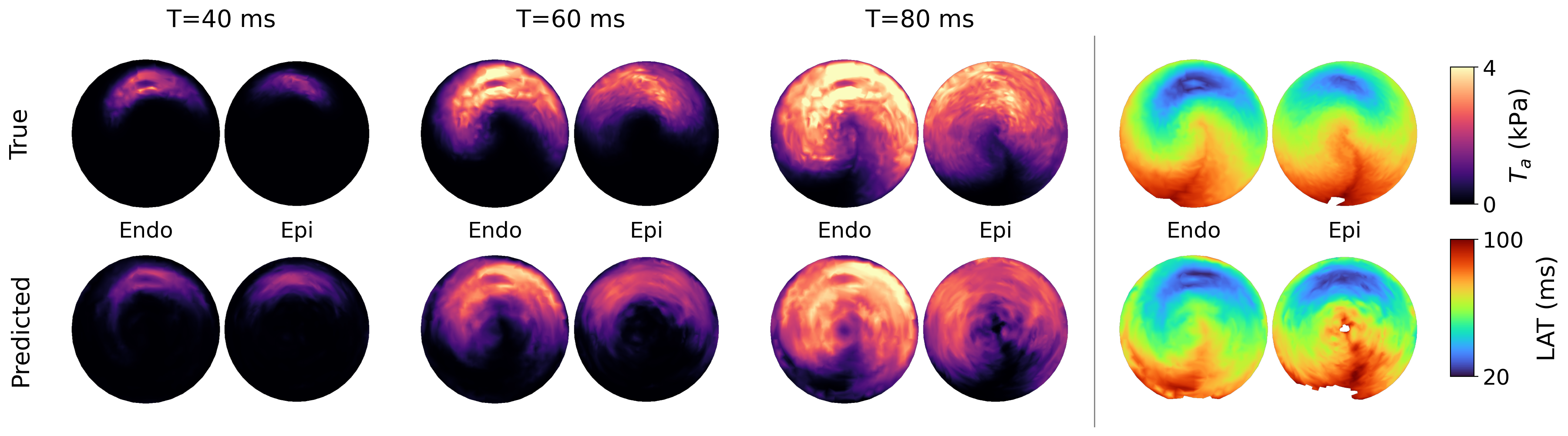}
        \caption{}
        \label{fig:reference-endo-epi-a}
    \end{subfigure}
    
    \begin{subfigure}{1.0\textwidth}
        \includegraphics[width=\linewidth]{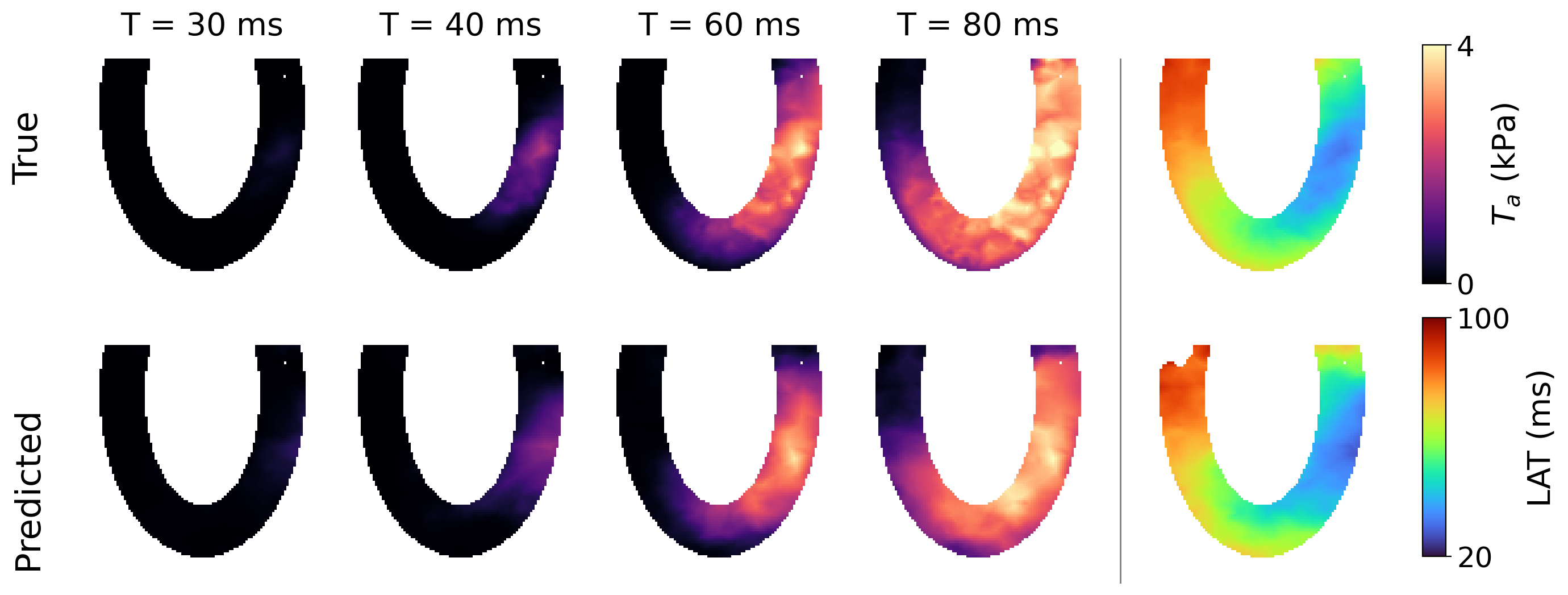}
        \caption{}
        \label{fig:reference-endo-epi-b}
    \end{subfigure}
    
    \caption[Endo- and epicardial polar maps for reference data]{PINN results for the estimated active tension $T_a$, optimized on all available data points. Panel \textbf{(a)} shows polar maps of the endo- and epicardial surfaces, calculated from the longitudinal and circumferential coordinates. The continuous field is visualized by linearly interpolating the vertex values in the new space. Panel \textbf{(b)} presents a cross-section of the volume by slicing the 3D volumetric mesh with the $x=-5$\;mm plane which, unlike other figures, does not go through the source point, revealing the transmural differences in wave propagation.}
    \label{fig:reference-endo-epi}

\end{figure}

To investigate the transmural variations in $T_a$ and LAT, panel (a) of Fig. \ref{fig:reference-endo-epi} shows the endo- and epicardial surfaces in polar (bullseye) view, where the radial direction in the figure corresponds to the longitudinal axis. More specifically, the coordinates are given by $x'=r\cos(\theta)$ and $y' = r\sin(\theta)$, with $r$ denoting the scaled longitudinal coordinate and $\theta$ the circumferential angle. As before, we see good correspondence between the true and predicted fields, both in propagation of the wave and magnitude. From this perspective, it is clearly seen that the active tension reaches higher values on the endocardial surface, which is also recovered by the PINN. Due to the transmural change in fiber direction, small differences in propagation direction are observed in the dataset, for example at time $T=40$\;ms. The PINN prediction captures these variations only partially, with the directional change appearing less pronounced. Panel (b) of Fig. \ref{fig:reference-endo-epi} illustrates these results as well. Here, the visualized cross-sections, unlike previous figures, do not intersect the source point, but are situated on the plane at $x=-5$\;mm  to the left of the source point in the surface plots of panel (a). Around $T=40$\;ms the wavefront has arrived first on the epicardium above the source, while the the opposite is true in the region below. The PINN prediction finds primarily the first discrepancy, resulting in an LAT map that shares the same diagonal line above the source but is less pronounced in the lower part. To recover the full transmural differences with highly varying fiber directions, two to three elements in the transmural direction might not be sufficient. 


\subsection{Effect of Gaussian noise}

\begin{figure}[p!]
    \centering
    
    \begin{subfigure}{1.0\textwidth}
        \includegraphics[width=\linewidth]{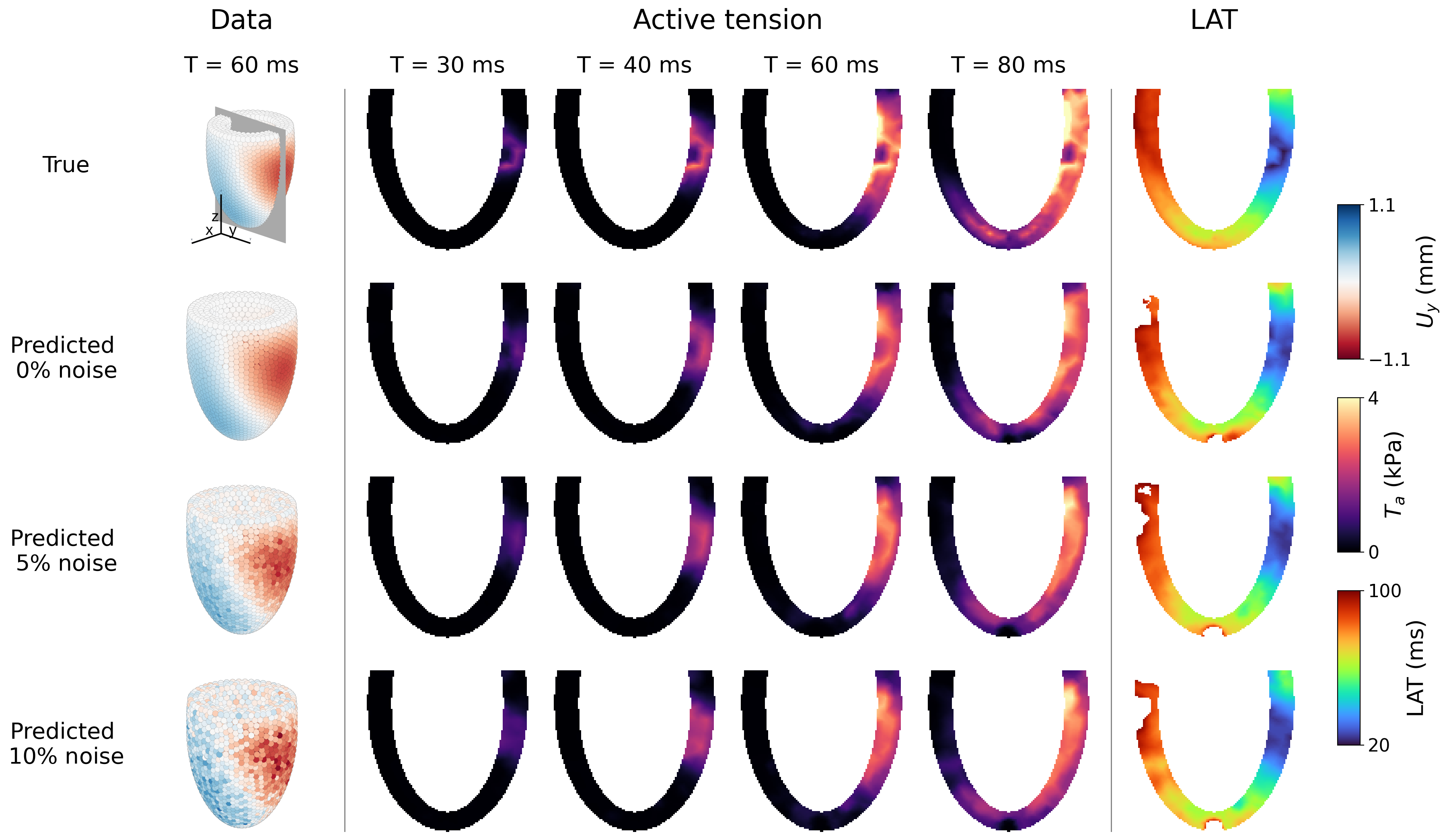}
        \caption{}
        \label{fig:noise-a}
    \end{subfigure}
    
    \begin{subfigure}{1.0\textwidth}
        \includegraphics[width=\linewidth]{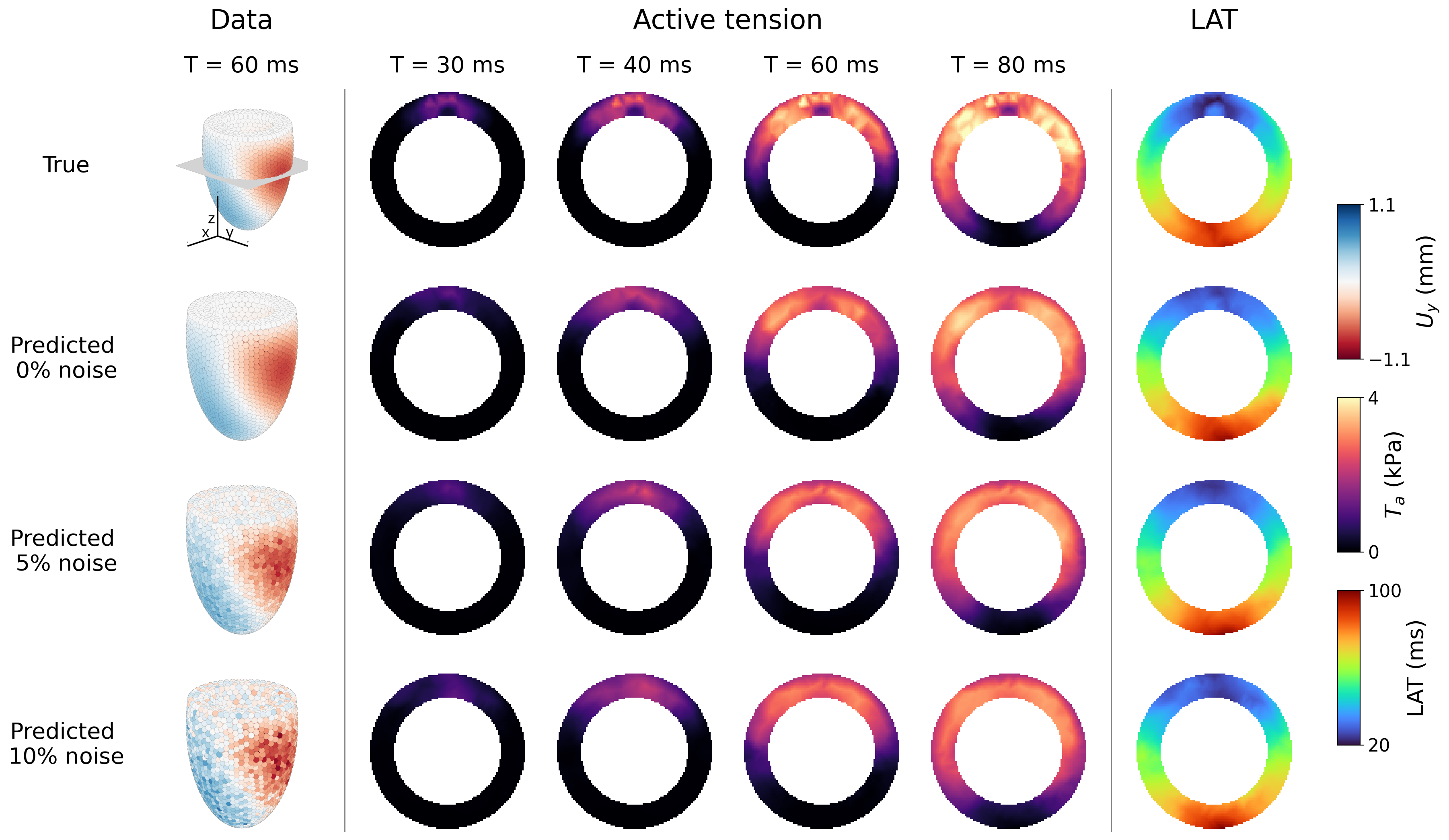}
        \caption{}
        \label{fig:noise-b}
    \end{subfigure}
    
    \caption[3D PINN results for Gaussian noise data]{PINN results for the estimated active tension $T_a$ and corresponding LAT map, optimized on all available data points with Gaussian noise levels of $5\;\%$ and $10\;\%$ of the maximum values. The datasets are illustrated in the first column by the y-component of the deformation at one timestep. The plane in the first row indicates the 3D orientation of the cross-sections, panel (\textbf{a}) for the longitudinal view (($x=0$\;mm) and panel (\textbf{b}) for the radial ($z=-5$\;mm). Both planes go through the source point.}
    \label{fig:noise}

\end{figure}

In this section, we investigate the reconstruction of the active tension wave when the deformation data is corrupted with noise. We added Gaussian variation on top of the deformation components with standard deviations equal to $5\;\%$ and $10\;\%$ of the maximum magnitude over all time steps, see Section \ref{ss:data-generation} for details. Fig. \ref{fig:noise} shows the effect on the estimated fields and compares the three datasets to the true fields. Panel (a) and panel (b) show the cross-sections through the source point for the longitudinal ($x=0$\;mm) and radial ($z=-5$\;mm) plane, respectively, similar to the reference figures (Figs. \ref{fig:reference-u-hp}, \ref{fig:reference-Ta}). The first column illustrates the Gaussian noise through a 3D scatter plot of the y-component of the deformation, as well as the cross-section plane used in the adjacent figures. Comparing the PINN predictions to the true solution, we can observe only a small influence of the added noise. The $T_a$-wave propagation is still clearly seen in the reconstructions, which is also reflected in the predicted LAT maps. Table \ref{tab:rmse}, which reports the RMSE across all variables, confirms that the added noise has only a minor impact on overall accuracy. Only features on fine spatial scales such as the absence of early $T_a$ at the source, diminish with increasing noise levels, resulting in slightly smoother fields. This can be clearly seen in panel (a) between the $0\;\%$ and $5\;\%$ case. Overall, we can conclude that the PINN demonstrates robustness to additive noise, while providing accurate reconstructions of wave propagation.

\subsection{Effect of reduced spatial resolution}

\begin{figure}[p!]
    \centering
    
    \begin{subfigure}{1.0\textwidth}
        \includegraphics[width=\linewidth]{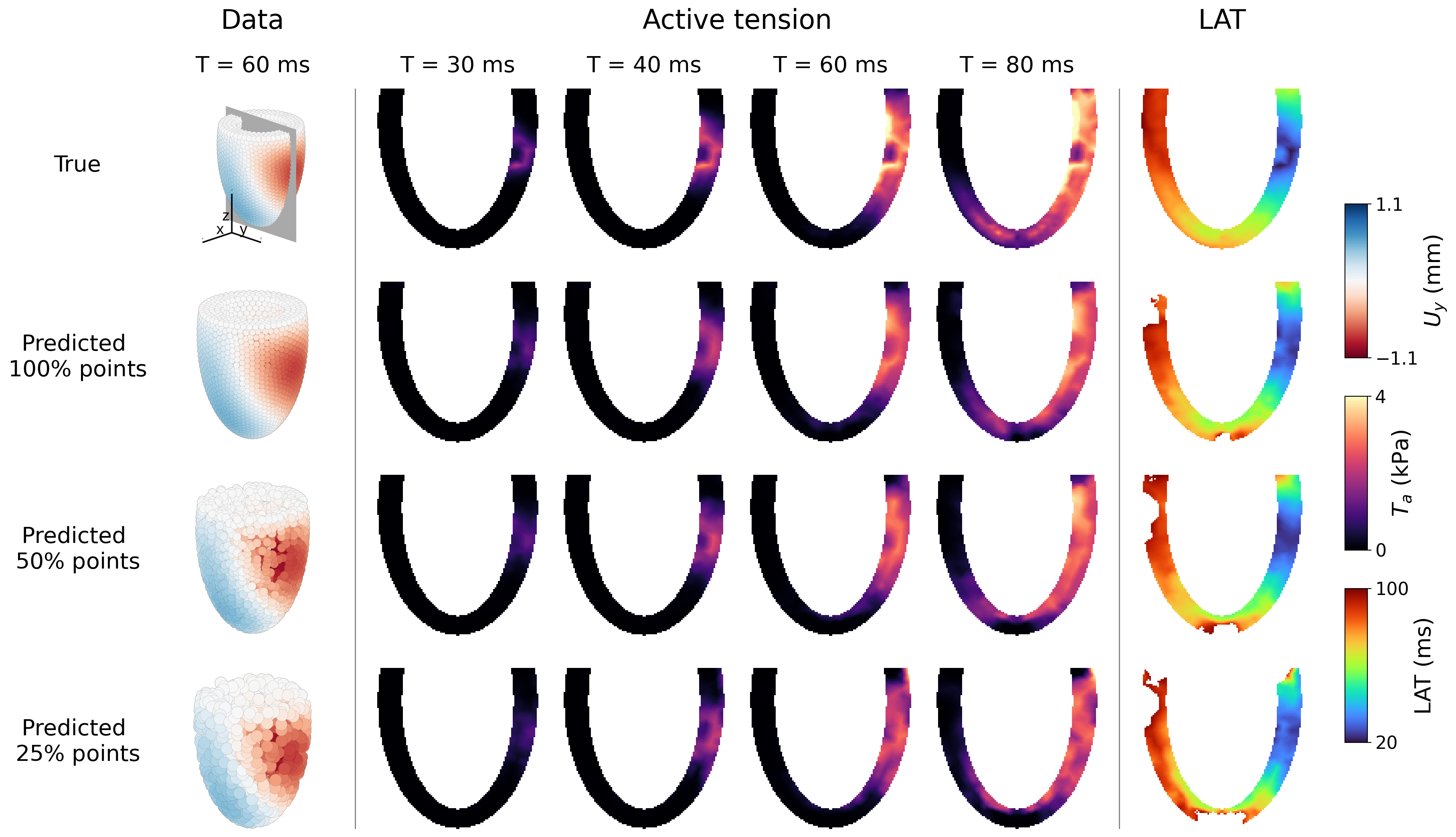}
        \caption{}
        \label{fig:resolution-a}
    \end{subfigure}
    
    \begin{subfigure}{1.0\textwidth}
        \includegraphics[width=\linewidth]{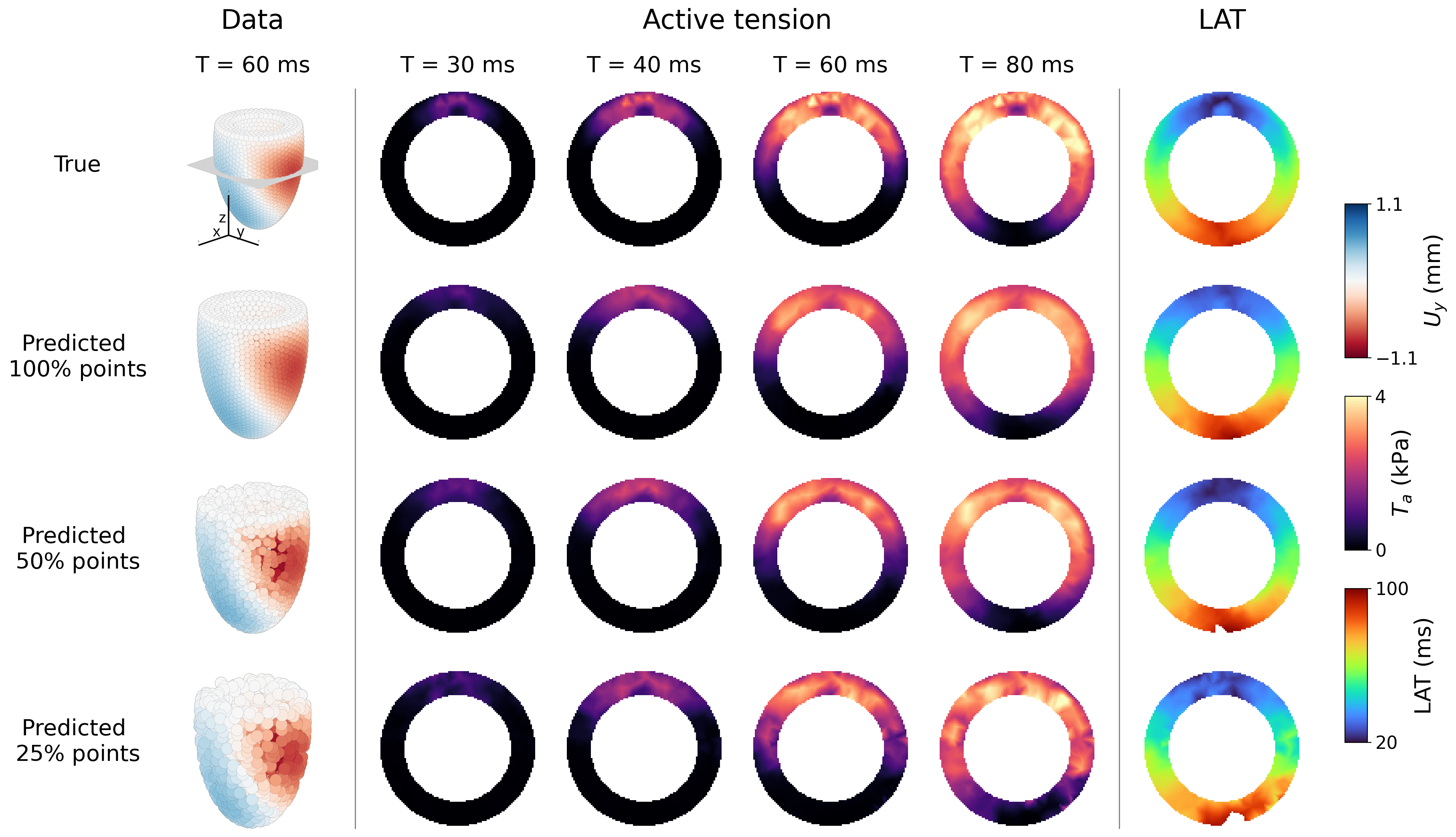}
        \caption{}
        \label{fig:resolution-b}
    \end{subfigure}
    
    \caption[3D PINN results for reduced resolution data]{PINN results for the estimated active tension $T_a$ and corresponding LAT map, optimized on data with reduced spatial resolution. The data was first smoothed with a discrete Gaussian before randomly subsampled to mimic reduced resolution in the data. This was done for $50\;\%$ and $25\;\%$ of all points, see Section \ref{ss:data-generation} for details. The datasets are illustrated in the first column by the y-component of the deformation at one timestep. The plane in the first row indicates the 3D orientation of the cross-sections, panel (\textbf{a}) for the longitudinal view (($x=0$\;mm) and panel (\textbf{b}) for the radial ($z=-5$\;mm). Both planes go through the source point.}
    \label{fig:resolution}

\end{figure}

Lastly, we tested the PINN framework against data with reduced spatial resolution. To this purpose, the original deformation data were smoothed with a Gaussian kernel and a subset of the available points was randomly selected. In Fig. \ref{fig:resolution} we present the results of using only $50\;\%$ and $25\;\%$ of the total dataset. Similar to the structure of Fig. \ref{fig:noise}, the used datasets are illustrated in the first column for the y-component of the deformation in which the size of the scatter points relates to the approximated volume they represent. Overall, PINN predictions optimized on lower-resolution datasets still follow the reference PINN estimation and the true field, as seen in both the $T_a$-field and the LAT map. However, unlike the $10\;\%$ noise case, accuracy of the prediction begins to decline after $T=60$\;ms declines when the resolution is reduced to $25\;\%$ of the total points. More specifically, the predicted $T_a$-fields seem coarser with steeper gradients, as shown in the longitudinal view of panel (a). Similarly, while the radial view in panel (b) shows that the magnitude of $T_a$ in the $25\%$ dataset approaches the true field at $T=80$\;ms, the overall propagation remains less accurate, as reflected in the LAT‑map. This loss of accuracy is also evident in Table \ref{tab:rmse}, where the RMSE of $T_a$ and the error in $p$ increase markedly, in contrast to the deformation error. The difficulty in predicting $p$ and $T_a$ could be partially explained by the lack of transmural information as the $25\%$ subsample dataset corresponds to a Gaussian smoothing with $\sigma=2/3$ of the average vertex distance. This heavily reduces the amount of information, especially in the transmural direction where only $3-4$ data points or $2-3$ elements are present in the original dataset. Generally speaking however, our proposed PINN framework is still able to estimate the active tension wave and related LAT map, even when data is reduced in resolution.

\begin{table}[ht!]
    \centering
    \resizebox{0.9\columnwidth}{!}{
    \begin{tabular}{|c|c|c|c|c|}
    \hline
    \multicolumn{1}{|c|}{Dataset} & \multicolumn{1}{c|}{$\alpha$} & \multicolumn{1}{c|}{RMSE $T_a$ (kPa)} & \multicolumn{1}{c|}{RMSE $U$ (mm)} & \multicolumn{1}{c|}{RMSE $p$ (kPa)}\\
    \hline
    \hline
    Reference & $10^{-1}$ & $0.333 \pm0.013$ &$(705.3\pm4.6)\cdot10^{-5}$  & $0.1811\pm0.0031$ \\ 
    \hline
    5$\%$ noise & $10^{0}$ & $0.3445\pm0.0045$ &$0.0127\pm0.0014$  & $0.2161\pm0.0025$ \\
    10$\%$ noise & $10^{0}$ & $0.388\pm0.028$ &$0.0156\pm0.0018$  & $0.2204\pm0.0053$ \\
    \hline
    0.50 resolution & $10^{-1}$ & $0.400\pm0.048$ &$0.00863\pm0.00052$  & $0.2248\pm0.0024$ \\
    0.25 resolution & $10^{-2}$ & $0.491\pm0.023$ &$0.00130\pm0.00022$  & $0.2799\pm0.0058$ \\
    \hline
    \end{tabular}}
    \caption[Overview of the selected $\alpha$ values and RMSE of estimated fields]{\label{tab:rmse} Summarization of the chosen $\alpha$ values for each dataset and the resulting PINN accuracies for the active tension $T_a$, the deformation $U$ and pressure $p$. The optimization is repeated $3$ times where the initialization of the NN weights, the mini-batch sampling and the gaussian noise if applicable is different. Average values over the runs and standard deviation are shown. The RMSE value of one run is calculated by taking the average over the full spatial and temporal domain, while all three components of $U$ are averaged jointly.}
    \end{table}

\section{Discussion}\label{s:discussion}

In this work, we employed Physics-Informed Neural Networks to address the inverse problem of estimating active tension waves from observed deformation in a LV geometry. Building on our previous studies with 2D square domain and linear models \cite{dermul2024}, we extended the approach to more complex scenarios concerning both spatial domain and material model. The undeformed 3D domain was defined as a truncated ellipsoid, representing an idealized LV. Deformation was described using non-linear theory with incompressible, hyperelastic material properties and anisotropic passive behavior. Active stress generation was also anisotropic, producing stress only along the fiber direction, which varied significantly across the wall. To tackle this challenging inversion task, we adopted the recently proposed delta-PINN framework to simultaneously estimate 3D-deformation, active tension, and hydrostatic pressure fields. As shown in Fig. \ref{fig:reference-Ta}, the method accurately reconstructed active tension propagation, with activation times closely matching the ground truth. This was achieved despite the activation fields originating from detailed biophysically inspired cell models and the inherent data–model mismatch in the function space of $\mathbf{U}$. Fig. \ref{fig:reference-endo-epi} further demonstrated that transmural variations were captured in the reconstruction, although the true differences between endocardial and epicardial surfaces were not fully reproduced. This discrepancy could stem from the larger $T_a$-magnitudes and the smaller spatial dimensions of the endocardial surface, which, together with the limited number of elements across the wall thickness, tend to bias the reconstruction in favor of the endocardium. To assess robustness, we tested the method on data of reduced quality. With Gaussian noise levels up to 10$\%$ of the total deformation, the reconstructions remained consistent, losing only small, finer spatial features. This confirmed that PINNs can effectively handle noisy data and that incorporating deformation reconstruction within the framework is feasible. We also examined the impact of reduced spatial resolution by smoothing and subsampling the data, thereby heavily diminishing the original information. Even under these conditions, the estimated $T_a$ preserved global propagation patterns in both active tension and LAT maps. However, in the most extreme case of retaining only 25$\;\%$ of the points, reconstruction accuracy declined, which was also seen in the estimated pressure fields.\\

The delta-PINN framework is built upon several principles. A central aspect of the framework lies in its input space, which is constructed from the eigenfunctions of the Laplacian. Owing to spectral bias often seen in PINNs \cite{rafiq2022,taylor2023}, these eigenfunctions provide a powerful means of representing the domain and, as demonstrated, are capable of capturing complex spatial patterns. To account for variations in the transmural direction, we included the first 200 Laplacian eigenfunctions. However, this number could potentially be reduced through more targeted identification of transmural modes, selectively incorporating them into the input space rather than relying on a broad inclusion of all preceding modes. In addition, the time dimension was introduced as one input node alongside the spatial eigenfunctions. Although this leads to imbalance in the input space, the network could still be optimized successfully. Nevertheless, alternative strategies, such as transforming temporal coordinates or coupling them directly to spatial coordinates, could offer improved efficiency. Another important property is the choice to optimize a weak formulation of the governing equations, explored in other PINN studies as well \cite{li2021,Gao2022}. In the original Delta-PINN paper \cite{SahliCostabal2024}, examples were presented using the energy formulation, where minimizing the energy density function of hyperelastic materials proved efficient. This approach avoids explicit computation of stress tensors and relies on a relatively simple energy expression (see Eq. \ref{eq:guccione}). However, in our preliminary tests for this problem, direct minimization of the energy did not yield a well-posed inverse problem for estimating $T_a$. This limitation, also noted in \cite{Herrmann2024}, led us to adopt the full Galerkin method with more extensive physical loss formulations. \\

The finite element formulation of the loss function, defined over the mesh nodes, introduced further advantages. For instance, Dirichlet boundary conditions could now be enforced strictly instead of incorporating them softly through an additional loss term. Moreover, restricting the search space to nodal values simplifies the optimization problem. The large amount of FEM research can also help the development of this kind of discrete PINNs. For example, we observed a recurring difficulty in estimating hydrostatic pressure, particularly at the boundaries. In principle, sufficient information should exist to recover two scalar fields from three balance equations, given that deformation is partially known. However, similar challenges have been known to exist in forward mixed FEM formulations where the inf-sup condition is not fully satisfied. One possible way to address this limitation is to employ second-order elements for the deformation field, although the associated increase in computational cost is not ideal. Alternatively, a quasi-incompressibility approach could be adopted, in which a large bulk modulus penalizes volumetric changes and removes the need to explicitly estimate $p$. A further strategy would be to decouple the deformation gradient and stress tensor into isochoric and volumetric components. All these strategies are possible in principle, but their relative performance and stability will likely depend on the specific data and desired accuracy.\\

To advance this work toward clinical applications, further research will be required to validate the methodology. A first step is to investigate how parameter mismatches affect the final solution. Some parameters could be estimated simultaneously, while others must be selected a priori from literature values or incorporated as patient-specific data to ensure individualization of the model. The choice of hyperparameters within the PINN framework is also important. In this study, we set $\alpha$ to values that broadly balanced the loss terms and prevented instability during optimization. However, more advanced balancing schemes could be introduced to improve efficiency \cite{xiang2022,mcclenny2023}. Such refinements, combined with more effective GPU parallelization, would enable faster and more stable optimization. Both speed and flexibility are important properties as they represent central advantages of PINNs over large supervised machine learning models, which, although accurate, are inherently more rigid and require extensive retraining when conditions change. 

Second, we have not yet investigated the reconstruction of the transmembrane potential directly, instead relying on active tension as a proxy. In abnormal or arrhythmic conditions, these quantities may diverge, necessitating additional inversion strategies. Incorporating more realistic boundary conditions, such as endocardial blood pressure, will also be essential. This can be achieved naturally within the weak formulation by introducing different Neumann boundary conditions and a single additional scalar parameter to be optimized. 

Finally, validation on more realistic synthetic datasets will be necessary, ideally generated through a full simulation pipeline combined with echocardiographic measurements, capturing observational noise and resolution effects more accurately. Such testing will help determine whether additional regularization of the active tension fields is required. The temporal dimension, in particular, offers a promising avenue for introducing this regularization, as it is not yet fully exploited in the current framework. Other studies \cite{dermul2024,Hofler2025}, for example, have imposed fixed temporal schemes, either during optimization or in a subsequent step, thereby reducing the estimation problem to a smaller set of parameters rather than requiring reconstruction across the entire timeline.

With regard to clinical potential and impact, physics-informed approaches have the advantage that they can impose physiological constraints, diminishing the risk of hallucinations or non-physical solutions seen in supervised methods. Such constraints also add to the explainability and robustness of the method. Also, since the PINN is optimized to reconstruct activation in a single patient, optimization is computationally relatively cheap, and no data needs to be distributed between hospitals. This approach therefore also does not require data augmentation using computer simulations, which introduces methodological concerns and biases in supervised machine learning methods applied to small datasets. 

Conceptually, PINNs offer a manner to integrate artificial intelligence (AI) approaches in medicine in a trustworthy manner, as this type of reconstruction lies closer to medical imaging than pure data analysis seen in supervised methods. 

The pathway for application to patient data is also clear. High-framerate ultrasound imaging can capture with one non-invasive imaging modality the deforming geometry of the heart, replacing the simplified LV shape in this work. Fitting the geometry in moving meshes will benefit from modern reconstruction operating with sparse data, which is also possible with PINNs \cite{pmlr-v250-verhulsdonk24a}. We expect a significantly better reconstruction compared to e.g. ECGi in the interventricular septum, which is not directly accessible via body-surface potentials. Validation of this method can be performed via non-invasive prediction of ectopic sources in the ventricle, pacemaker lead positions or infarcts, and validated via endocardial mapping or magnetic resonance imaging. The advantage of the ultrasound methodology is that it contains no ionizing radiation, and its probes are relatively cheap relative to medical imaging hardware. When combined with non-invasive ablation methods \cite{cuculich_noninvasive_2017}, this research line contributes to fully external treatment of cardiac rhythm disturbances. 



\section{Conclusion}\label{s:conclusion}

In this study, we demonstrated that physics‑informed neural networks can effectively reconstruct activation waves in three‑dimensional left ventricular geometries. The non-convex 3D geometry, passive material models, and anisotropy required a more advanced PINN framework, which incorporated transformed spatial inputs, weak formulations, and finite element–based loss functions. As a result, we were able to accurately reconstruct active tension waves alongside the deformation fields and hydrostatic pressure. The method proved robust to noise and reduced spatial resolution, preserving global propagation patterns and confirming its potential for handling imperfect clinical data. Looking forward, further refinement of parameter estimation, boundary conditions, and temporal regularization will be key steps toward clinical translation. With these advances, PINNs could provide a powerful tool for patient-specific modeling and non-invasive assessment of the electrical wave patterns in practice.

\section*{Declarations}


\begin{itemize}
\item This research was funded by KU Leuven project C24E/21/031 and the LUMC Fellowship 2024. 
\item The authors declare no conflict of interest. 
\item Ethics approval and consent to participate: not applicable.
\item Consent for publication: not applicable.
\item Data will be made publicly available on zenodo.org upon publication.
\item Materials availability: not applicable. 
\item Code will be made publicly available on gitlab.com upon publication. 
\item CRediT Author contributions: ND contributed to conceptualization, methodology, software and visualization and writing - original draft; HD contributed to conceptualization, funding acquisition, development of methodology and writing - reviewing and editing. 
\end{itemize}

\bibliography{3D_localcopy}

\end{document}